\documentclass[onecolumn,amsmath,amssymb,pra]{revtex4-2}

\usepackage{graphicx}
\usepackage{dcolumn}
\usepackage{amsmath}
\usepackage{CJK}
\usepackage{subfigure}
\usepackage{sidecap}
\usepackage{dsfont}

\usepackage{mathrsfs}
\usepackage{amsfonts}
\usepackage{indentfirst}
\usepackage{bm}

\usepackage[dvipsnames]{xcolor}

\usepackage[colorlinks,citecolor=ProcessBlue]{hyperref}

\newcommand{\be}{\begin{equation}}
\newcommand{\ee}{\end{equation}}
\newcommand{\bey}{\begin{eqnarray}}
\newcommand{\eey}{\end{eqnarray}}
\newcommand{\bw}{\begin{widetext}}
\newcommand{\ew}{\end{widetext}}

\newcommand{\ra}{\rangle}
\newcommand{\la}{\langle}

\newcommand{\ba}{\begin{array}}
\newcommand{\ea}{\end{array}}
\newcommand{\bi}{\begin{itemize}}
\newcommand{\ei}{\end{itemize}}
\newcommand{\bem}{\begin{enumerate}}
\newcommand{\eem}{\end{enumerate}}

\begin{document}

\title{Role of long-range interaction in critical quantum metrology}

\author{Zhen-Xia Niu$^{1}$}\email{niuzhx@zjnu.edu.cn}
\author{Qian Wang$^{1,2}$}\email{qwang@zjnu.edu.cn}

\affiliation{$^{1}$Department of Physics, Zhejiang Normal University, Jinhua 321004, China \\
$^2$CAMTP-Center for Applied Mathematics and Theoretical Physics, University of Maribor,
Mladinska 3, SI-2000, Maribor, Slovenia}

\date{\today}

\begin{abstract}

Long-range interacting quantum systems are useful for improving the performance of 
various applications of quantum technologies. 
In this work, we carry out a detailed analysis of 
how the long-range interaction affects the measurement precision 
in critical quantum metrology.
By employing the paradigmatic model of a Kiteav chain 
with power-law decaying interaction, we focus on the impacts of
long-range interaction on the critical sensing for the scenarios with and without
uncertainty in system parameters.
We show that the long-range interaction can be used as a valuable
resource for enhancing the sensitivity of critical parameter estimation in both scenarios.
Our findings not only provide more insights into the features 
of the long-range interacting systems, but also
verify the usefulness of long-range interacting systems in quantum metrology.

\end{abstract}

\maketitle

\section{introduction} 

Investigating how to precisely measure a set of unknown physical parameters
is important in both theoretical studies and various practical applications.
In this context, much attention has been devoted to improving measurement
precision by leveraging quantum features. 
This has been promoted the development of the so-called quantum metrology
\cite{Paris2009,Degen2017,Braun2018,Giovannetti2006,Giovannetti2011}
for achieving high precision measurements in 
terms of the quantum Fisher information (QFI),  
which quantifies the estimation precision 
via the quantum Cram\'er-Rao bound \cite{Braunstein1994}.
A key property of quantum metrology is the ability to surpass 
the ultimate limits of measurement precision imposed by 
classical physics \cite{Degen2017}.
This has been dubbed as the quantum enhanced sensitivity and
indicates the superiority of quantum metrology in the studies of parameter estimation.

The quantum enhanced sensitivity can be achieved by different schemes 
\cite{Braun2018,Montenegro2024}.
A well-known scheme is to exploit nonclassical quantum correlations, 
such as quantum entanglement and quantum squeezing, to improve the sensitivity. 
This was first verified by the pioneer work in Ref.~\cite{Giovannetti2004} 
and has received considerable experimental and theoretical 
interests over the past two decades, see a recent 
review \cite{Huang2024} and references therein.
However, due to the inherent fragility of the quantum entanglement \cite{Horodecki2009}, 
a practical implementation of this scheme is very challenging, if not impossible. 
It is therefore worthwhile to find other quantum enhanced measurement schemes 
that are more accessible to experiments.
A more promising scheme is the critical quantum metrology 
\cite{Ivanov2013,Tsang2013,Bina2016,Zanardi2008,Salvatori2014}, 
which has been extensively studied in recent years
\cite{Garbe2020,Mirkhalaf2021,Garbe2022b,Candia2023,
Rams2018,Frerot2018,Hotter2024,ChuY2021,Giovanni2022,Mihailescu2024,
Garbe2022a,Gietka2022,Ilias2022,Salvia2023,Ostermann2024,Alushi2024,ZhangR2024,Wald2020}
and is the focus of the present work. 

Instead of employing the highly entangled states to improve the measurement precision,
the sensitivity in critical quantum metrology is enhanced by the divergence 
of the quantum fluctuations in the proximity of the critical point.
The usefulness of quantum criticality for enhancing sensitivity has been demonstrated
in second- and first-order 
\cite{Zanardi2008,Salvatori2014,Garbe2020,Mirkhalaf2021,Garbe2022b,Candia2023,
Montenegro2021,Gammelmark2011,Skotiniotis2015,Sarkar2024,YangL2019,Mirkhalaf2020}, 
topological \cite{Sarkar2022,YangY2024,Mukhopadhyay2024}, 
Stark and quasiperiodic localization 
\cite{HeX2023,Rozhin2024,Sarkar2024b,Sahoo2024,LiangE2024} 
phase transitions, as well as in non-Hermitian systems \cite{Wiersig2020}. 
Apart from equilibrium quantum phase transitions, the quantum enhanced sensitivity
has also been observed in nonequilibrium quantum phase transitions, such as
dynamical \cite{GuanQ2021,ZhouL2023},
Floquet \cite{LangJ2015,Mishra2021,Mishra2022,ZhangK2023,Rodzinka2024}, 
dissipative \cite{Fernandez2017,Raghunandan2018,Heugel2019,Ivanov2020,Guillaume2024}, 
and time crystal \cite{Montenegro2023,Iemini2024,Cabot2024,Leo2024,RozhinY2024,Shukla2024} 
phase transitions. 
Moreover, the enhancement effect of quantum criticalities on the 
measurement precision has been experimentally verified using 
different platforms \cite{LiuR2021,DingD2022,Ilias2024}.

As quantum phase transitions usually occur in many-body systems with complex interactions,
an essential ingredient that leading to the enhancement of sensitivity 
in critical quantum metrology is the interaction between the system components.
On the one hand, the ability of interactions in quantum many-body systems 
for achieving high precision measurements has been 
revealed in both theoretical \cite{Boixo2007,Boixo2008}
and experimental  \cite{Napolitano2011,HouZ2021} studies. 
On the other hand, the features of quantum criticalities 
are strongly dependent on the interaction range.
It is known that the long-range interactions can alter the universal 
property of quantum critical systems \cite{Defenu2023}.
The presence of long-range interactions also results in various unpercedented phenomena,
including anomalous thermalization \cite{VanR2016}, 
novel dynamical phase transitions \cite{Defenu2019,Halimeh2020}, 
and metastable phase \cite{NDefenu2021,Giachetti2023},
see Refs.~\cite{Defenu2023,Defenu2024} for a thoroughly review.
These facts motivate us to explore the role played by long-range interaction in
critical quantum metrology. 
Although the impacts of the long-range interaction on the dynamical quantum metrology 
have been investigated in previous work \cite{YangJ2022},     
a detailed understanding of its effects on critical quantum metrology remains unknow.

In the present work, we carry out our exploration in 
long-range Kitaev (LRK) chain \cite{Vodola2014,Maity2019},
which is a prototypical model in the studies of long-rang interacting quantum systems.
The LRK chain undergoes quantum phase transition for both short- 
and long-range interaction cases \cite{Vodola2014,Dutta2018},
making it a suitable model for our purpose.
Moreover, it is an analytically tractable model, enabling us 
to explicitly analyze the imprints of long-range interaction 
in critical quantum metrology.
We investigate how the performance of critical quantum metrology responds
to the variation of interaction range in two different scenarios.
The first one is the common single parameter estimation, 
in which only the concerned parameter is unknown 
while all other parameters are perfectly known. 
Another one is the uncertain critical quantum metrology \cite{Mihailescu2024},
which allows some uncertainties presented in the parameters 
that are required to be precisely known in the single parameter estimation. 
We show that the long-range interaction
can be utilized as a useful resource to achieve 
high sensitivity parameter estimations in both scenarios.   

The rest of this article is structured as follows.
In Sec.~\ref{Second}, we briefly review the general 
setting of quantum parameter estimation.
The basic features of our considered system, namely the LRK chain, 
are introduced in Sec.~\ref{Third}.
We evaluate the QFI and report our results in Sec.~\ref{Fourth}. 
Specifically, in Sec.~\ref{FrA}, we show the role of 
long-range interaction in the critical quantum metrology 
for the scenario of single parameter estimation.
The effects of long-range interaction on the measurement precision 
for the uncertain critical quantum metrology are analyzed in Sec.~\ref{FrB}.
We finally conclude our studies in Sec.~\ref{Fiv} with several remarks.

\section{Quantum parameter estimation} \label{Second}

A general scenario of quantum estimation is to estimate a set of unknown
parameters $\mathbf{x}$ by performing suitable measurements 
on a quantum probe $\rho(\mathbf{x})$, which encodes 
the information of the unknown parameters 
\cite{Giovannetti2006, Giovannetti2011, Liu2020}.
The estimated values of the unknown parameters are inferred
through an estimator, which is a function of the measurement outcomes. 
The precision achievable for jointly estimating parameters $x_a$ is
constrained by the quantum Cram\'er-Rao bound \cite{Braunstein1994,Paris2009,Liu2020}
\be
   Cov(\mathbf{x})\geq\frac{1}{\mathcal{N}}\mathcal{F}^{-1},
\ee
where $\mathcal{N}$ is the number of measurement repetitions, and
$Cov(\mathbf{x})$ denotes the covariance matrix with elements,
$Cov(x_a,x_b)=\la(x_a-\la x_a\ra)(x_b-\la x_b\ra)\ra$. 
Here, $\mathcal{F}$ is the QFI matrix with entries
\be \label{EntryQFIM}
   \mathcal{F}_{ab}=2\sum_{ij}\frac{\mathrm{Re}
            [\la i|\partial_a\rho|j\ra\la j|\partial_b\rho|i\ra]}{p_i+p_j}.
\ee
where $\partial_a:=\partial/\partial x_a$ and the probe state has been decomposed 
as $\rho(\mathbf{x})=\sum_ip_i|i\ra\la i|$.
The diagonal element $\mathcal{F}_{aa}$ quantifies the ultimate precision limit 
of the single parameter estimation, in which the unknown parameter $x_a$ is estimated 
and all other parameters are precisely known.  

It has been proved that the quantum Cram\'er-Rao bound can be saturated
if and only if $\mathrm{Tr}(\rho[L_a,L_b])=0$.
Here, $L_a$ is the symmetric logarithmic derivative (SLD) operator 
with respect to the parameter $x_a$ and defined as
$\partial_a\rho=(\rho L_a+L_a\rho)/2$.
Using the SLD operators for parameters $\mathbf{x}$, the 
elements of QFIM can be rewritten as
$\mathcal{F}_{ab}=\mathrm{Tr}(\rho\{L_a,L_b\})/2$
with $\{\cdot\}$ being the anti-commutation. 
One can easily find that, the SLD operators 
for pure states $\rho=|\psi\ra\la\psi|$ are $L_a=2\partial_a\rho$.
As a result, $\mathcal{F}_{ab}$ in (\ref{EntryQFIM}) can be simplified to
\be
  \mathcal{F}_{ab}=4\mathrm{Re}[\la\partial_a\psi|\partial_b\psi\ra-
              \la\partial_a\psi|\psi\ra\la\psi|\partial_b\psi\ra].
\ee
This expression of $\mathcal{F}_{ab}$ shows that QFI of pure states is determined by the
change rate of the probe state with respect to the estimated parameters.
As quantum systems are highly susceptible to the variation 
of the driving parameters in the proximity of their critical points
\cite{Campos2007,Zanardi2007,GuJ2010},
the quantum criticality can be used as a resource to enhance measurement precision, 
leading to the so-called critical quantum metrology
\cite{Frerot2018,Hotter2024,ChuY2021,Giovanni2022,Garbe2022b,Bayat2024}.

To achieve the ultimate precision bound defined by QFI, the measurement 
procedure should be performed in the optimal basis.
Among different choices of the optimal basis, 
a well-known one is the projectors composed 
by the eigenstates of the SLD operator \cite{Paris2009,Meyer2021}.
However, as the optimal measurements for different parameters are noncommutative, 
it is more complex to saturate the Cram\'er-Rao bound 
in multiparameter metrology \cite{Albarelli2020}.

\section{Long-rang Kitaev model} \label{Third}

The Hamiltonian of the long-range Kitaev (LRK) model
can be written as \cite{Vodola2014,Dutta2018,Defenu2023}
\begin{align} \label{LRKH}
 H(\mu)=&-\frac{t}{2}\sum_{j=1}^L\left(f_j^\dag f_{j+1}+f_{j+1}^\dag f_j\right)
    -\mu\sum_{j=1}^L\left(f_j^\dag f_j-\frac{1}{2}\right)
   +\frac{\Delta}{2}\sum_{j=1}^L\sum_{y=1}^{L-j}
   d_y^{-\alpha}\left(f_j f_{j+y}+f_{j+y}^\dag f_j^\dag\right),
\end{align}
where $f_j$ ($f_j^\dag$) is the fermionic annihilation (creation) operator for site $j$
and $L$ denotes the total number of sites.
Here, the antiperiodic boundary condition $f_j=-f_{j+L}$ has been assumed. 
The parameter $t$ is the hopping strength, $\mu$ is the on-site chemical potential, and
$\Delta$ represents the pairing strength that is characterized by an algebraic decay 
with distance $d_y=\mathrm{min}(y, L-y)$ and exponent $\alpha$.
In this work, without loss of generality, we set $\Delta=1$.

 \begin{figure}
  \includegraphics[width=\columnwidth]{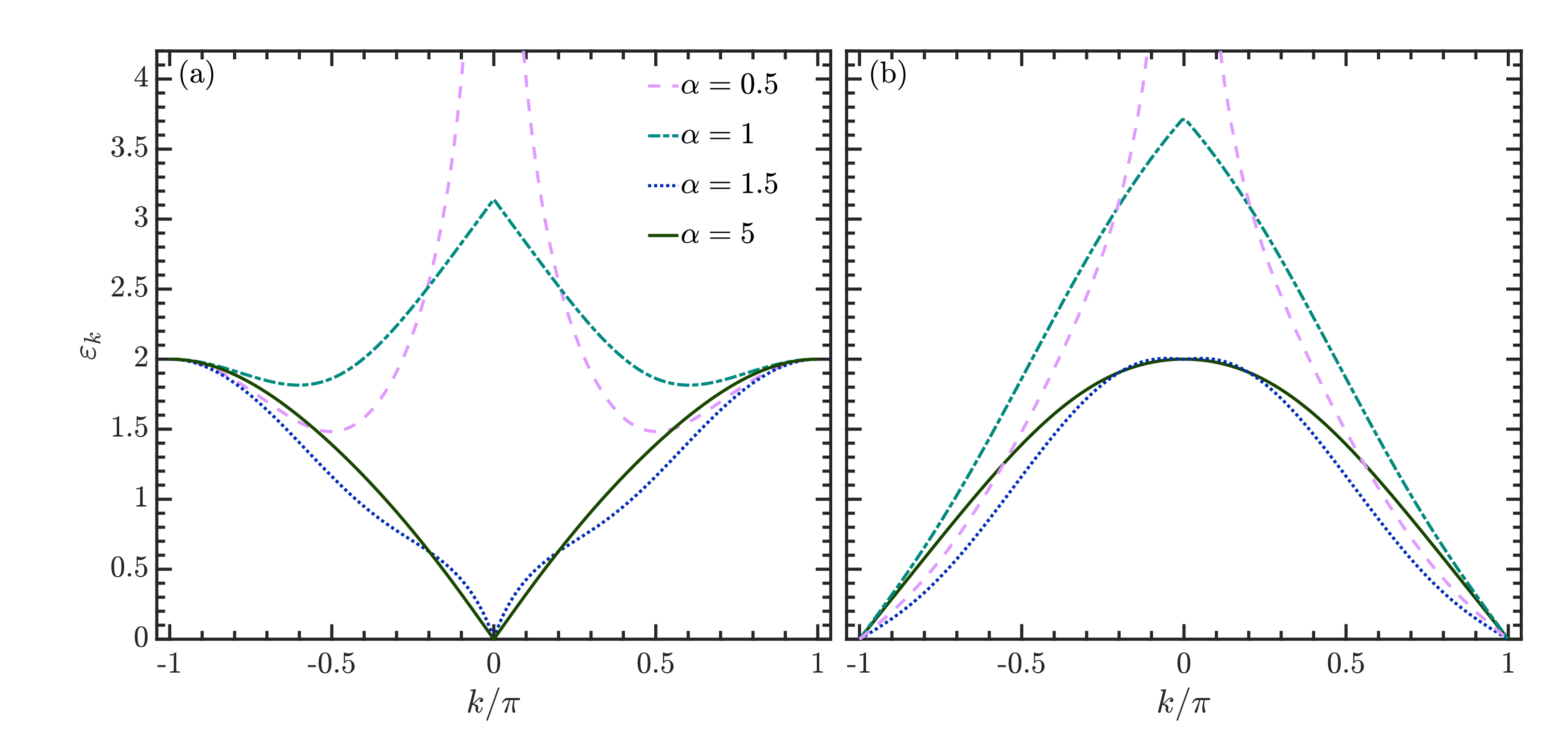}
  \caption{Quasiparticle dispersion $\varepsilon_k$ of LRK model 
   as a function of $k$ with (a) $\mu=-t=-1$ and 
  (b) $\mu=t=1$ for several $\alpha$ values.
  Other parameter: $\Delta=1$.}
  \label{Edk}
 \end{figure}

As the Hamiltonian (\ref{LRKH}) is quadratic in fermions, it can be
diagonalized as follows.
First, by using the Fourier transform
$f_j=\frac{e^{-i\pi/4}}{\sqrt{L}}\sum_ke^{-ik j}c_k$, with
$k=\frac{(2n+1)\pi}{L}$ and $n=-L/2,\ldots,L/2-1$ for even $L$ and
$n=-(L-1)/2,\ldots,(L-1)/2$ for odd $L$,
we recast the Hamiltonian (\ref{LRKH}) as 
$H(\mu)=\sum_{k>0}\mathcal{C}_{k}^\dag\mathcal{H}_{k}(\mu)\mathcal{C}_{k}$.
Here, $\mathcal{C}_{k}^\dag=(c_{k}^\dag, c_{-k})$ is the Nambu spinor and
\be
 \mathcal{H}_{k}(\mu)=
 \begin{bmatrix}
  -(\mu+t\cos k)  & f_\alpha(k)  \\
  f_\alpha(k) & (\mu+t\cos k)
 \end{bmatrix},
\ee
with $f_\alpha(k)=\sum_{y=1}^{L-1}\sin(ky)/d_y^\alpha$ 
being the Fourier transform of the long-range interaction term 
and encoding the long-range nature of the system.
Then, by performing a Bogoliubov transformation
$c_{k}=\cos(\theta_k)\gamma_k+\sin(\theta_k)\gamma_{-k}^\dag$
with $\tan(2\theta_{k})=-f_\alpha(k)/(\mu+t\cos k)$, one can find that
Hamiltonian (\ref{LRKH}) is finally diagonalized as
$H(\mu)=\sum_{k>0}\varepsilon_k
(\gamma_{k}^\dag\gamma_{k}+\gamma_{-k}^\dag\gamma_{-k}-1)$
with $\varepsilon_k=\sqrt{(\mu+t\cos k)^2+[f_\alpha(k)]^2}$
being the quasiparticle dispersion \cite{Vodola2014,VanR2016,Dutta2017,Francica2022}.
It is also easy to find that the ground state of $H$ (\ref{LRKH}) is given by
\be \label{GST}
   |\Psi_0\ra=\prod_{k>0}\left(\cos\theta_k+\sin\theta_k c_k^\dag c_{-k}^\dag\right)|0\ra,
\ee
where $|0\ra$ denotes the vacume state of $c_{\pm k}$ fermions.

In the thermodynamic limit $L\to\infty$, the function $f_\alpha(k)$ can be written in terms 
of the polylogarithmic function which 
vanishes at $k=0,\pm\pi$ for all $\alpha>1$ and diverges as $k^{\alpha-1}$ 
in the limit $k\to0$ for $\alpha<1$ \cite{Olver2010}.
As a result, the dispersion $\varepsilon_k$
is gapless for $|\mu|=t$ when $\alpha>1$, indicating the presence of two 
quantum phase transitions at $\mu/t=\pm 1$ \cite{Vodola2014}.
In contrast, even though the criticality at $\mu=t$ is preserved for $\alpha<1$, 
there is no phase transition at $\mu=-t$ in this case \cite{Vodola2014,Baghran2024}.

In Figs.~\ref{Edk}(a) and \ref{Edk}(b), we show how $\varepsilon_k$ varies 
as a function of $k$ for different values of $\alpha$ 
with $\mu/t=\pm 1$, respectively.
For $\mu/t=-1$, the quasiparticle dispersion $\varepsilon_k$ 
vanishes at $k=0$ when $\alpha>1$, while $\varepsilon_k>0$ as long as $\alpha\leq1$,
as observed in Fig.~\ref{Edk}(a). 
Hence, the critical point at $\mu/t=-1$ for $\alpha>1$ disappears once $\alpha\leq1$.
However, as shown in Fig.~\ref{Edk}(b), 
$\varepsilon_k$ is always zero at $\mu/t=1$ for $k=\pm\pi$,
regardless of the value of $\alpha$. 
This means that the criticality at $\mu/t=1$ is independent of the $\alpha$ value.

In the rest of this work, we focus on the role played by the interaction range 
in quantum critical metrology \cite{Frerot2018,Garbe2020,Hotter2024}. 
As the critical point $\mu/t=1$ is robust to the change of the interaction range, 
we restrict our study to the case of $\mu/t\geq0$ and consider only $\alpha>1$ case. 
By studying the behavior of the QFI, 
we aim to reveal how the interaction range affects the precision 
of parameter estimation without and with the 
uncertainties \cite{Mihailescu2024} in the system's control parameters.

 \begin{figure}
  \includegraphics[width=\columnwidth]{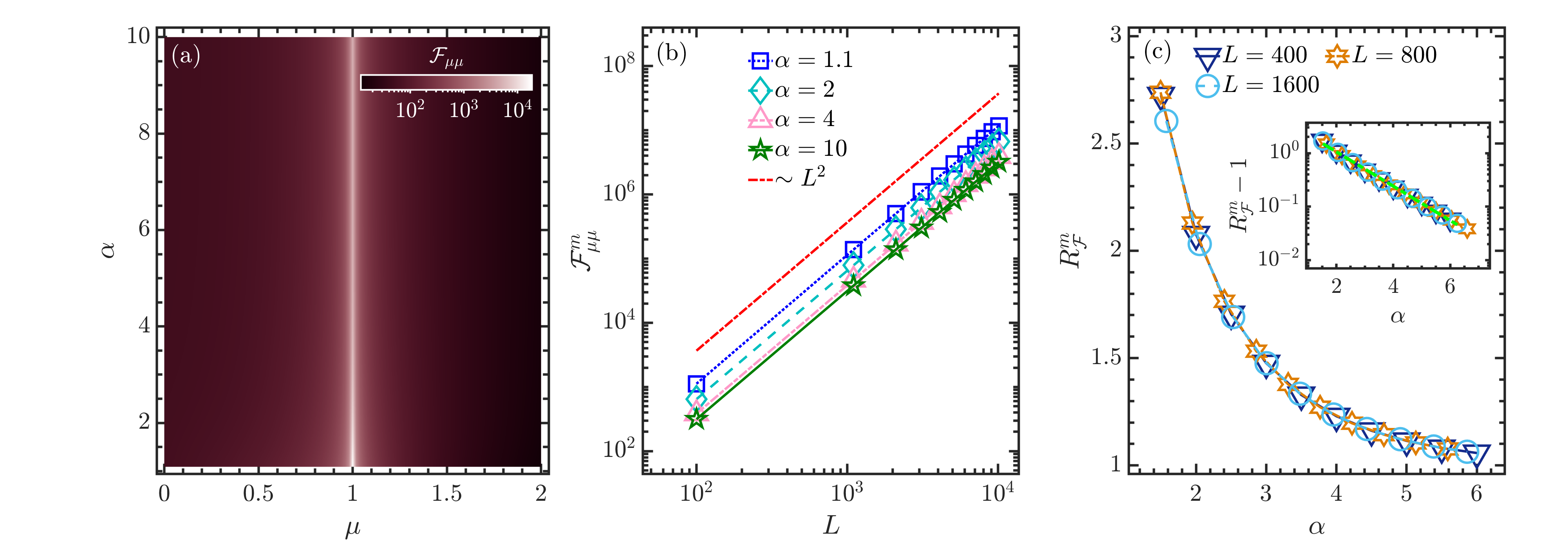}
  \caption{(a) QFI $\mathcal{F}_{\mu\mu}$ as a function of $\mu$ and $\alpha$ with the system 
  size $L=400$.
  (b) Scaling of the maximal value of QFI, $\mathcal{F}_{\mu\mu}^m$, 
  as a function of the system size $L$
  for different values of $\alpha$. 
  The red dot-dashed line denotes the power law of the form
  $\mathcal{F}_{\mu\mu}^m\propto L^2$. 
  (c) $R_\mathcal{F}^m$ in Eq.~(\ref{Rfm}) as a function of $\alpha$ for several system sizes.
  The inset plots the variation of $R_\mathcal{F}^m-1$ with increasing $\alpha$.
  The green dot-dashed line shows the scaling of $R_\mathcal{F}^m$ with the interaction range
  $\alpha$, $R_\mathcal{F}^m-1\sim e^{-p\alpha}$ with $p\simeq0.7378$. 
  Other parameters: $t=\Delta=1$.}
  \label{Esmu}
 \end{figure}

\section{Results and discussions} \label{Fourth}

The ground state $|\Psi_0\ra$ in (\ref{GST}) is a pure and tensor product state.
Hence, the additive feature of QFI allows us to
decompose the QFI matrix of $|\Psi_0\ra$ as 
$\mathcal{F}=\mathcal{F}_{ab}(|\Psi_0\ra)=\sum_{k>0}\mathcal{F}_{ab}(|\psi_0^k\ra)$,
where $|\psi_0^k\ra=\cos\theta_k|0_k,0_{-k}\ra+\sin\theta_k|1_k,1_{-k}\ra$
with $|n_k,n_{-k}\ra (n_{\pm k}=0,1)$ being the eigenstates of $c_{\pm k}^\dag c_{\pm k}$.
For each $k$, the QFIM of $|\psi_0^k\ra$ is given by
\be
  \mathcal{F}_k=\mathcal{F}_{ab}(|\psi_0^k\ra)=4\mathscr{R}
         \left[\la\partial_a\psi_0^k|\partial_b\psi_0^k\ra
           -\la\partial_a\psi_0^k|\psi_0^k\ra\la\psi_0^k|\partial_b\psi_0^k\ra\right]
           =4\partial_a\theta_k\partial_b\theta_k
\ee
where $\mathscr{R}(z)$ denotes the real part of $z\in\mathbb{C}$.

In our study, we set $a=t, b=\mu$, then the explicit form 
of $\mathcal{F}_{ab}(|\psi_0^k\ra)$ can be written as 
\be
\mathcal{F}_k=\mathcal{F}_{ab}(|\psi_0^k\ra)=
\begin{pmatrix}
\dfrac{f_\alpha^2(k)\cos^2 k}{[(\mu+t\cos k)^2+f_\alpha^2(k)]^2},\ 
\dfrac{f_\alpha^2(k)\cos k}{[(\mu+t\cos k)^2+f_\alpha^2(k)]^2} \\
\dfrac{f_\alpha^2(k)\cos k}{[(\mu+t\cos k)^2+f_\alpha^2(k)]^2},\ 
\dfrac{f^2_\alpha(k)}{[(\mu+t\cos k)^2+f_\alpha^2(k)]^2}
\end{pmatrix},
\ee
and thus $\mathcal{F}=\sum_k\mathcal{F}_k$.
Note that $\mathrm{det}[\mathcal{F}]=\mathrm{det}[\sum_k\mathcal{F}_k]=0$
for all $L$ and $\alpha$, indicating the dependence 
between $\mu$ and $t$ \cite{Mihailescu2024,Goldberg2021}.
This is verified by the fact that the critical point is 
determined by $\mu/t$, rather than these two parameters separately.

\subsection{Role of $\alpha$ in single parameter estimation} \label{FrA}

Let us first investigate the impacts of the interaction range $\alpha$ 
on the estimation of $\mu$ with all
other parameters are precisely known.
In this case, the attainable precision of $\mu$ is given by 
$\delta\mu\geq[\mathcal{F}_{\mu\mu}]^{-1}$, with
\be \label{QFImu}
  \mathcal{F}_{\mu\mu}=\sum_k\mathcal{F}_{\mu\mu}^k
     =\sum_k\frac{f^2_\alpha(k)}{[(\mu+t\cos k)^2+f^2_\alpha(k)]^2}.
\ee
Clearly, the QFI of $\mu$ with finite $\alpha$ value is more complex than 
the $\alpha=\infty$ case, in which $f_\alpha(k)$ is simplified into $f_\infty(k)=\sin k$.  

In Fig.~\ref{Esmu}(a), we show how the QFI $\mathcal{F}_{\mu\mu}$ varies as
a function of $\mu$ and $\alpha$ with $L=400$ and $t=1$.
As the ground state of a many-body system is 
very sensitive to the change of parameters near the critical point,
we see that $\mathcal{F}_{\mu\mu}$ exhibits a sharp peak around the critical point,
regardless of the $\alpha$ value. 
This means that the advantage of the critical metrology still holds for the critical systems
with long-range interactions.
To uncover the critical feature of $\mathcal{F}_{\mu\mu}$, we first note that
its behavior near the critical point $\mu=1$ is governed by the modes with $k\approx k_c=\pi$.
Hence, we can rewrite $\cos k=-1+(\pi-k)^2/2+O((\pi-k)^2)$ 
and expand $f_\alpha(k)$ as follows \cite{Solfanelli2023}
\begin{align}
   f_\alpha(k)&=(1-2^{2-\alpha})\frac{\zeta(\alpha-1)}{\zeta(\alpha)}(\pi-k)+O((\pi-k)^3),
   \quad \alpha>1\ \text{and}\ \alpha\neq2, \\
   f_\alpha(k)&=\frac{6}{\pi^2}[2\ln2-1-\ln(\pi-k)](\pi-k)+O((\pi-k)^3),\quad \alpha=2, \label{Ap2}
\end{align}
where $\zeta(x)$ is the Riemann zeta function.
Then, inserting these expansions in Eq.~(\ref{QFImu}) with $t=1$ 
and taking $\mu\to1, L\gg1$, one can find that maximal value of QFI, 
denoted by $\mathcal{F}_{\mu\mu}^m$, has following scaling behavior
\be \label{FmaxS}
  \mathcal{F}^m_{\mu\mu}\propto\frac{1}{(\pi-k)^2}=\frac{L^2}{\pi^2},
\ee
where $k=k_{max}=\pi-\pi/L\approx\pi$ has been used to get the final equality.
Here, we would like to point out that the logarithmic term in the expansion of $f_k(\alpha)$
for $\alpha=2$ case [cf.~Eq.~(\ref{Ap2})] gives rise to subleading 
contributions to the scaling of $\mathcal{F}_{\mu\mu}^m\sim L^2$. 
However, we can omit this subleading contributions in the larger $L$ limit.

 \begin{figure}
  \includegraphics[width=\columnwidth]{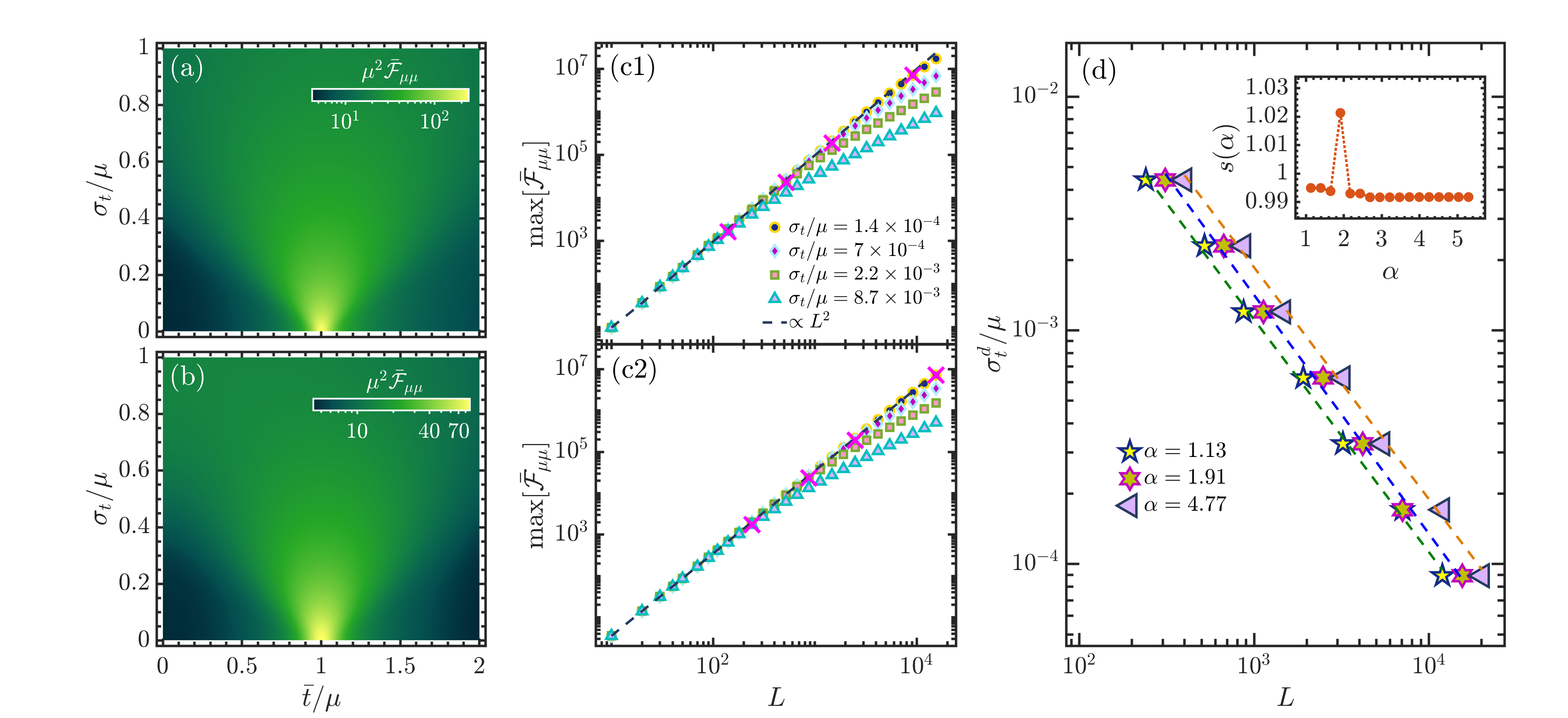}
  \caption{(a)-(b): Attainable precision of $\mu$ with 
  uncertainty in $t$ varies as a function of the degree of
  uncertainty $\sigma_t/\mu$ and $\bar{t}/\mu$ for $\alpha=1.3$ 
  (a) and $\alpha=5$ (b) with the system size $L=50$. 
  (c1) and (c2) plot the maximum QFI, $\mathrm{max}[\bar{\mathcal{F}}_{\mu\mu}]$,
  as a function of the system size $L$ for several degrees of uncertainty in $t$ with $\alpha=1.3$ (c1)
  and $\alpha=5$ (c2).
  The dark blue dashed line denotes the power law scaling
  $\mathrm{max}[\mathcal{F}_{\mu\mu}]\propto L^2$ for the case 
  when $t$ is perfectly known.
  For each $\sigma_t/\mu$ value, the magenta cross marks the point
  at which the scaling of
  $\mathrm{max}[\bar{\mathcal{F}}_{\mu\mu}]$ no longer obeys the power law form. 
  (d) $\sigma_t^d/\mu$, defined as the degree of uncertainty at which the dependence of
  $\mathrm{max}[\bar{\mathcal{F}}_{\mu\mu}]$ on the system size deviates from the 
  power law behavior, as a function of $L$ for several $\alpha$ values.
  The dashed lines represent a power-law fit $\sigma_t^d/\mu\sim L^{-s(\alpha)}$
  with $\alpha$-dependent power-law exponent $s(\alpha)$. 
  The inset of panel (d) shows how $s(\alpha)$ varies as a function of $\alpha$. 
  Other parameter: $\Delta=1$.}
  \label{AvgQFI}
 \end{figure}

The result in (\ref{FmaxS}) indicates that the scaling of $\mathcal{F}_{\mu\mu}^m$
is independent of the interaction range and attains the Heisenberg limit \cite{Giovannetti2004}.
This confirms that the quantum criticality is a useful resource for quantum metrology. 
The variation of $\mathcal{F}_{\mu\mu}^m$ with system size $L$ 
for different values of $\alpha$ are plotted in Fig.~\ref{Esmu}(b).
One can clearly see that the behavior of $\mathcal{F}_{\mu\mu}^m$ is well 
captured by the scaling in (\ref{FmaxS}), irrespective of the $\alpha$ value.
Moreover, the results in Fig.~\ref{Esmu}(b) also show that $\mathcal{F}_{\mu\mu}^m$
decreases with increasing the range of interaction, implying that the 
long-range interaction can boost measurement precision of critical quantum metrology.

The role played by the interaction range in the
critical quantum estimation can be revealed by the ratio between the maximal QFI of 
finite- and short-range interacting systems, and defined as
\be \label{Rfm}
  R_\mathcal{F}^m=\frac{\mathcal{F}_{\mu\mu}^m}{\mathcal{F}_{\mu\mu}^m(\infty)}.
\ee
Here, $\mathcal{F}_{\mu\mu}^m$ and $\mathcal{F}_{\mu\mu}^m(\infty)$ 
are, respectively, the maximal value of $\mathcal{F}_{\mu\mu}$ in (\ref{QFImu})
for the cases of finite $\alpha$ and $\alpha=\infty$.
According to the definition of $R_\mathcal{F}^m$, we have $R_\mathcal{F}^m>1$
if the long-range interaction is enable to enhance the precision of the critical quantum metrology.   
On the contrary, $R_\mathcal{F}^m<1$ implies that 
the long-range interaction has detrimental effect 
on the quantum estimation as compared to its short-range counterpart. 

In Fig.~\ref{Esmu}(c), we plot $R_\mathcal{F}^m$ as a function of $\alpha$
for several system sizes.
We see that $R_\mathcal{F}^m>1$ and decreases with increasing the $\alpha$ value,
regardless of the system size.
This means that a substantial measurement advantage can be provided by the long-range
interacting quantum systems.
Moreover, the scaling behavior of $\mathcal{F}_{\mu\mu}^m$ in (\ref{FmaxS})
leads to the same dependence of $R_\mathcal{F}^m$ 
on the interaction range $\alpha$ for different system sizes. 
Another feature that can be obviously observed in Fig.~\ref{Esmu}(c) is
the convergence of $R_\mathcal{F}^m$ to one in the limit $\alpha\to\infty$.
This is consistence with the definition of $R_\mathcal{F}^m$ in Eq.~(\ref{Rfm}).
In particular, we find that the decrease of $R_\mathcal{F}^m$ with increasing $\alpha$
is well described by an exponent decay of the form
$R_\mathcal{F}^m-1\sim e^{-p\alpha}$,
with $\alpha>1$ and $p\simeq0.7378$, as seen in the inset of Fig.~\ref{Esmu}(c). 

These results demonstrate that the long-range interaction is an useful resource in 
critical quantum metrology.
However, above discussions have assumed that except the parameter of interest, all other 
parameters must be perfectly known.
This is a rather stringent requirement in practice, and, particularly, 
for the quantum critical systems.
Hence, it is necessary to consider the uncertainty in control parameters for
studying of critical sensing.
Motivated by this fact, the uncertain quantum critical metrology has been proposed 
in a very recent work \cite{Mihailescu2024}, in which the effect of the
uncertainty of system parameters on the precision of quantum estimation
has been investigated in different critical systems.
In the following of this section, we explore whether above unveiled long-range advantage
still holds in uncertain quantum critical sensing.

\subsection{Role of $\alpha$ in uncertain critical quantum estimation} \label{FrB}

In general, the ground state of a quantum many-body system is determined by many
control parameters at the critical point.
The presence of uncertainty in any control parameters other than the probed one
would lead to detrimental effect on the performance of critical estimation. 
The uncertain quantum critical estimation aims to evaluate
the disadvantage of the uncertainties in system parameters on metrological utility.  
To this end, it employs appropriate probability distribution to characterize 
the uncertainty of a certain control parameter \cite{Mihailescu2024}. 

In the present work, we consider the sensitivity of $\mu$ 
when the parameter $t$ is known with some uncertainty.
Our goal is to investigate whether the long-range interaction can also enhance
the measurement precision of $\mu$ for different uncertainties in $t$.
The presence of uncertainty in $t$ implies that the probe state 
no longer be the ground state with certain $t$ value. 
However, it becomes an infinite mixture of ground states 
that correspond to different choice of $t$.
Hence, the probe state can be constructed as
\be \label{MixGS}
   \rho(\mu,\sigma_t)=\int_{-\infty}^\infty p(t)\rho_g(\mu,t)dt,
\ee
where $\rho_g(\mu,t)=|\psi_g\ra\la\psi_g|$ is the density matrix 
of the ground state for a given $t$ value and each $|\psi_g\ra$ is obtained from
a Gaussian ensemble with probability 
$p(t)=e^{-(t-\bar{t})^2/2\sigma_t^2}/(\sigma_t\sqrt{2\pi})$.
Here, $\bar{t}$ denotes the average value of $t$ and $\sigma_t$
quantifies the degree of uncertainty in parameter $t$.
One can recover the single parameter estimation 
by setting $\sigma_t\to0$, which results in $p(t)\to\delta(t-\bar{t})$, meaning that
the parameter $t$ is precisely known.
Contrarily, the multi-parameters estimation corresponds to $\sigma_t\to\infty$ case, 
in which $t$ is completely unknown.

 \begin{figure}
  \includegraphics[width=\columnwidth]{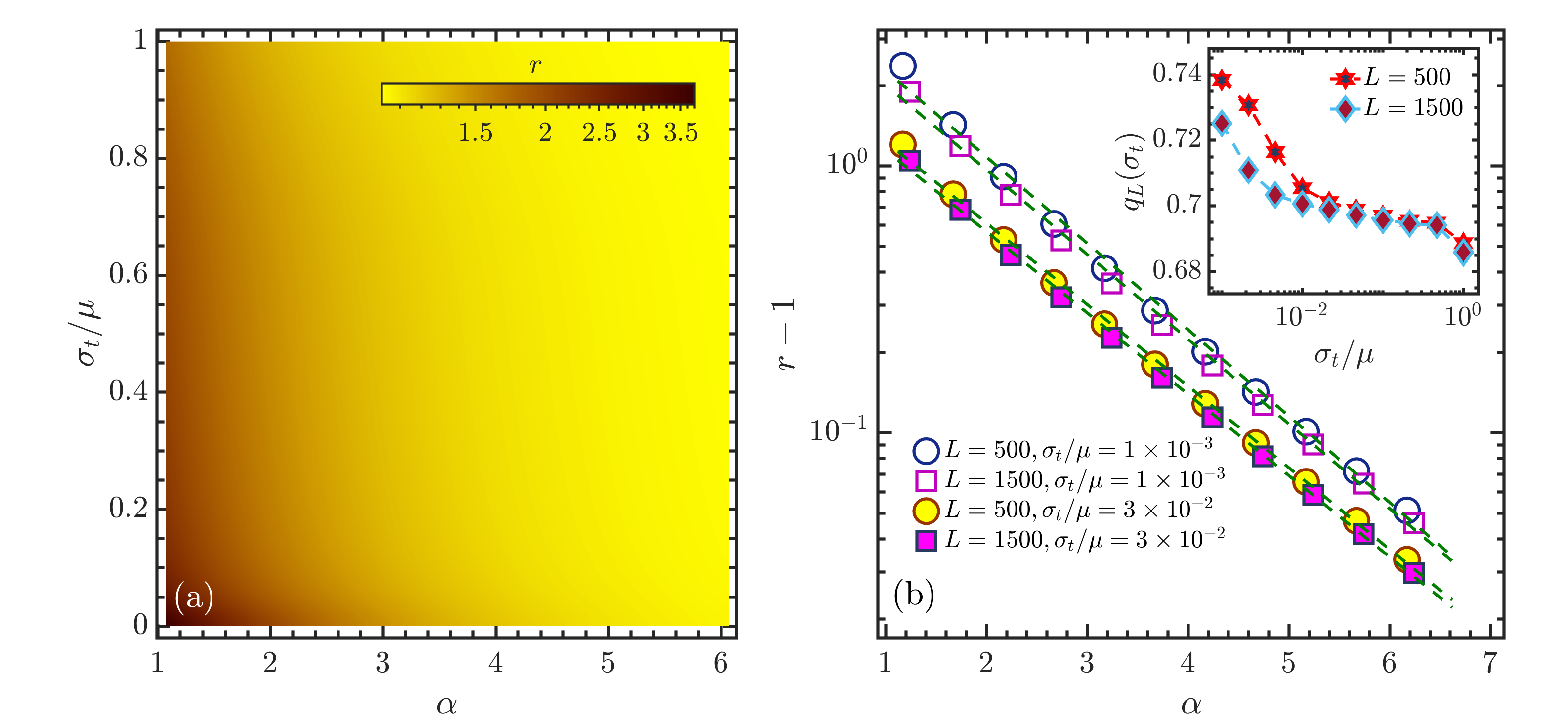}
  \caption{(a): $r$ in Eq.~(\ref{RMaxQFI}), defined as
  the ratio between $\mathrm{max}[\bar{\mathcal{F}}_{\mu\mu}]$ 
  and its short-range limit counterpart,
  as a function of $\alpha$ and $\sigma_t$ with system size $L=50$. 
  (b) $r-1$ as a function of $\alpha$ using line-log axes for several system sizes
  and $\sigma_t$ values.
  The dashed lines denote the exponential decay behavior of the form
  $r-1\sim e^{-q_L(\sigma_t)\alpha}$ with exponent $q_L(\sigma_t)$ depends on
  both $L$ and $\sigma_t$.
  The inset in panel (b) depicts $q_L(\sigma_t)$ as a function of $\sigma_t$
  for different system sizes. 
  Other parameter: $\Delta=1$.}
  \label{RatioQFI}
 \end{figure}

As $\rho(\mu,\sigma_t)$ in (\ref{MixGS}) can be considered as an average
over the ground states weighted probability $p(t)$,  
its QFI with respect with $\mu$,
denoted by $\bar{\mathcal{F}}_{\mu\mu}$, can be calculated as
\be
   \bar{\mathcal{F}}_{\mu\mu}=\int_{-\infty}^\infty dt p(t)\mathcal{F}_{\mu\mu},
\ee
where $\mathcal{F}_{\mu\mu}$ is given by Eq.~(\ref{Esmu}).
Clearly, $\bar{\mathcal{F}}_{\mu\mu}$ reduces to $\mathcal{F}_{\mu\mu}$ when $\sigma_t=0$,
while it decreases with increasing $\sigma_t$ for the parameter $t$.
This means that the critical enhanced sensitivity will be fading away as the degree of uncertainty
in the parameter $t$ is increased. 
To see this, we plot $\mu^2\bar{\mathcal{F}}_{\mu\mu}$ as a function of $\bar{t}/\mu$ and 
$\sigma_t/\mu$ for a system size $L=50$ with $\alpha=1.3$ and $\alpha=5$ 
in Figs.~\ref{AvgQFI}(a) and \ref{AvgQFI}(b), respectively.  
One can obviously observe that the behavior of $\bar{\mathcal{F}}_{\mu\mu}$ 
is similar for different $\alpha$ values.
Specifically, for small $\sigma_t/\mu$ values, the QFI
exhibits a peak near the critical point, such as in the single parameter case. 
However, the peak of $\bar{\mathcal{F}}_{\mu\mu}$ quickly 
disappear with increasing the degree of uncertainty in the parameter $t$. 
Moreover, one can further find that $\bar{\mathcal{F}}_{\mu\mu}$ turns into a
smooth function as $\sigma_t/\mu$ is increased. 
These features of the QFI are quite similar to the results 
observed in the Ising model \cite{Mihailescu2024},
which corresponds to the short-range limit with $\alpha=\infty$, 
Hence, regardless of the interaction range, 
the presence of uncertainty in any control parameters other than the concerned one could 
lead to the negative effect on the quantum critical estimation,
resulting in the loss of critical enhancement.

Although the critical sensitivity is dissolved when the uncertainty in $t$ is large, 
it retains for the cases with small $\sigma_t$ values.
It is therefore necessary to analyze whether the critical sensitivity of $\mu$ with uncertainty in $t$
still scales as in the single parameter scenario with $t$ is precisely known and, particularly,
what is the impact of the interaction range on the scaling behavior of the QFI.
In Figs.~\ref{AvgQFI}(c1) and \ref{AvgQFI}(c2), we plot the maximum QFI, 
$\mathrm{max}[\bar{\mathcal{F}}_{\mu\mu}]$, as a function of system size $L$ 
for several degrees of imprecision in $t$ with $\alpha=1.3$ and $\alpha=5$.
In order to highlight the imprint of uncertainty in $t$, we also plot the power-law scaling $L^2$
for the single parameter case with $t$ is exactly known as dark blue dashe lines in both figures.
One can clearly see that introducing the uncertainty in $t$ makes the 
scaling of the maximum QFI deviate from the power-law form.
It is also obvious that the degree of the deviation from the power-law scaling in the behavior of
$\mathrm{max}[\bar{\mathcal{F}}_{\mu\mu}]$ increases with increasing $\sigma_t$.
This indicates the detrimental effect of the imprecision in $t$ on the 
quantum critical metrology, in agreement with Ref.~\cite{Mihailescu2024}.
Importantly, we note that the degree of deviation of the maximum QFI 
from the $L^2$ scaling behavior is more apparent in the long-range interaction case,
as evidenced by the magenta crosses Figs.~\ref{AvgQFI}(c1) and \ref{AvgQFI}(c2).
Here, the magenta crosses are identified by the condition
$|1-\max[\bar{\mathcal{F}}_{\mu\mu}]/\mathcal{F}_{\mu\mu}^m|>\delta_d$,
marking the points at which the maximum QFI no longer follows the $L^2$ scaling behavior.
The threshold $\delta_d$ is an arbitrary small value.
In our studies, we choose $\delta_d=0.1$, but a careful numerical
check has been verified that our main results are 
independent from the choice of $\delta_d$ value.
These findings demonstrate that the long-range interacting
system is more susceptible to the presence of uncertainty in
control parameter than its short-range 
counterpart in uncertain quantum critical sensing.

To further investigate the effect of the interaction range 
on the uncertain quantum critical sensing, we study how the
uncertainty in $t$ corresponding to the magenta crosses,
denoted by $\sigma_t^d/\mu$,
varies as a function of system size $L$ for different interaction ranges.
Our results are shown in Fig.~\ref{AvgQFI}(d). 
It can be seen that $\sigma_t^d/\mu$ decreases with increasing $L$, 
regardless of the interaction range.
Hence, the larger the system size, the smaller the uncertainty of $t$
in order to keep the same precision as in the single parameter scenario.
Moreover, the best numerical fit reveals that the scaling of $\sigma_t^d/\mu$ 
with $L$ is well captured by $\sigma_t^d/\mu\sim L^{-s(\alpha)}$ with
$s(\alpha)$ being the $\alpha$-dependent exponent. 
The explicity dependence of $s(\alpha)$ on the interaction range $\alpha$ is 
plotted in the inset of Fig.~\ref{AvgQFI}(d).
We see clearly that $s(\alpha)$ exhibits a peak around $\alpha=2$.
As $\alpha=2$ corresponds to the boundary between the long- and short-range interaction
regimes, the peak in $s(\alpha)$ can be use to diagnose the presence of transition from
the long-range to short-range interaction, or vice versa.
Away from the $\alpha=2$ point, the behavior of $s(\alpha)$ is almost independent
of the interaction range in both $\alpha<2$ and $\alpha>2$ regimes.
However, it is crucial to note that the value of $s(\alpha)$ 
for long-range interactions ($\alpha<2$) is slightly larger 
than the short-range ($\alpha>2$) cases.
As a result, as compared to the long-range interacting systems, 
the short-range interactions are more robust with respect
to experimental imprecision in uncertain quantum critical metrology.

Another obvious feature exhibited by Figs.~\ref{AvgQFI}(a) and \ref{AvgQFI}(b) is that
the long-range interaction gives rise to larger QFI, irrespective of 
the uncertainty in the parameter $t$.
Hence, even though the long-range interaction makes the system become 
more sensitive to uncertainties, it can increase the QFI value for any degrees of uncertainty.
This fact leads us to claim that the long-range interaction could enhance
the sensitivity of an unknown concerned parameter with 
uncertainty in other system parameters.
To confirm this statement, as done in the single parameter paradigm, we 
compare $\mathrm{max}[\bar{\mathcal{F}}_{\mu\mu}]$ 
to its short-range limit counterpart, 
denoted by $\mathrm{max}[\bar{\mathcal{F}}_{\mu\mu}(\infty)]$, 
and define their ratio as
\be \label{RMaxQFI}
    r=\frac{\mathrm{max}[\bar{\mathcal{F}}_{\mu\mu}]}
        {\mathrm{max}[\bar{\mathcal{F}}_{\mu\mu}(\infty)]}. 
\ee
The results in Figs.~\ref{AvgQFI}(a) and \ref{AvgQFI}(b) imply that
$r$ is defined in the interval $r\geq1$. 
The advantage of long-range interaction in uncertain quantum critical metrology 
is manifested by $r>1$, while we have $r=1$ in the short-range limit $\alpha\to\infty$.

The variation of $r$ as a function of $\alpha$ and $\sigma_t$ 
is shown in Fig.~\ref{RatioQFI}(a).
One can clearly see that the long-range interaction leads to larger 
$\mathrm{max}[\bar{\mathcal{F}}_{\mu\mu}]$, regardless of
the degree of uncertainty in $t$.  
This verifies the usefulness of the long-range interaction for improving the
measurement precision in quantum critical metrology
even when the certain system parameter cannot be perfectly known.
We further examine how $r$ approaches $1$ as $\alpha$ is increased for different 
values of $\sigma_t$ and $L$.
The results are plotted in Fig.~\ref{RatioQFI}(b).
It is obvious that the decrease of $r-1$ with increasing $\alpha$ is well captured by
an exponential behavior of the form $r-1\sim e^{-q_L(\sigma_t)\alpha}$, irrespective
of the degree of uncertainty and the system size.
However, the decay exponent $q_L(\sigma_t)$ is decreased when 
we increase $\sigma_t$ or the system size $L$, as seen in the inset of Fig.~\ref{RatioQFI}(b).

\section{Conclusions} \label{Fiv}

In this work, we have scrutinized whether and how the interaction range affects
the performance of the quantum critical sensing in the long-range Kitaev model
with power-law decaying interaction, which exhibits quantum critical behavior
at certain control parameters.
The analytical feature of the Kitaev model enables us to exactly calculate the 
quantum Fisher information (QFI) matrix with respect to our estimated parameter. 
This in turn provides us more insight into the role played by interaction range
in quantum critical estimation.

We mainly focused on two scenarios. 
The first one is the single-parameter paradigm, 
in which, excepting the interested parameter,
all other system parameters are assumed precisely known.
For this scenario, we have shown that the QFI retains
its peak near the critical point as the interaction range is increased.
Moreover, the presence of long-range interaction does not change the super-linear scaling 
of the maximum QFI with system size, which satisfies the Heisenberg limit.
This means that the critical enhanced sensitivity with resepct to
the system size resource is unaffected by the long-range interaction.
However, we have found that the QFI increases exponentially with increasing the
range of interaction, indicating the usefulness of the long-range interaction as a resource
for improving critical sensitivity.

In practice, the complexity of the quantum critical many-body systems
hinders us to perfectly know all system parameters, implying that
the assumption in single parameter paradigm is too stringent to be implemented.  
Hence, a more realistic and appropriate scenario for quantum critical estimation 
should allow relevant experimental uncertainties in some control parameters,
resulting in the so-called uncertain quantum critical metrology \cite{Mihailescu2024}.
It is therefore necessary to investigate 
the role of interaction range in this second paradigm.
We have demonstrated that the system with long-range interaction is more
sensitivity to the uncertainty than the short-range interacting system.
Most importantly, we have verified that the long-range interaction still plays  
as a useful resource for critical sensing 
even when we introduce the imprecision in system parameters.

%%%This reflects the negative effect of the long-range interaction on the critical sensing
%%%in more practical scenario.  

Our findings imply potential applications of the long-range 
interacting systems in quantum sensing 
for achieving high precision measurements.
Although one can expect that the long-range advantage revealed in the present work
should be more prominent in strong long-range regime ($\alpha<1$),
a detailed examination is still required due to the different 
dispersion relation in this regime \cite{NDefenu2021,Giachetti2023}.
Another natural extension of the present work is 
to analyze the impact of long-range interaction on the critical quantum sensing
that bear different kinds of noise \cite{Garbe2020,Ostermann2024,Alushi2024}.
Finally, as the quantum long-range interacting systems
can be experimentally realized in trapped-ion platforms \cite{Defenu2023}, 
we expect that our studies could motivate further experimental 
explorations of long-range advantage in quantum sensing.

 \acknowledgments
  
Q.~W acknowledge the support from the Slovenian Research and Innovation Agency (ARIS)
under Grants No.~J1-4387 and No.~P1-0306.
Z.~X.~N acknowledges the financial support from Zhejiang Provinical Nature Science
Foundation under the Grant No.~LQ22A040006.
This work was also support from Zhejiang Provinical Nature Science
Foundation under the Grant No.~LY20A050001.

\bibliographystyle{apsrev4-2}
%\bibliography{QESLRC}

%apsrev4-2.bst 2019-01-14 (MD) hand-edited version of apsrev4-1.bst
%Control: key (0)
%Control: author (72) initials jnrlst
%Control: editor formatted (1) identically to author
%Control: production of article title (-1) disabled
%Control: page (0) single
%Control: year (1) truncated
%Control: production of eprint (0) enabled
%

\end{document}